\documentclass[final]{aipproc}
\layoutstyle{6x9}

\usepackage{graphicx}
\usepackage{amsmath,amssymb}

\newcommand{\rmd}{{\rm d}}
\newcommand{\rme}{{\rm e}}
\newcommand{\rmi}{{\rm i}}
\newcommand{\BesselI}{{\rm I}}

\begin{document}

\title{Two interacting electrons in a magnetic field: comparison of semiclassical, quantum, and variational solutions}


\author{Tobias Kramer}{
  address={Institut f\"ur Theoretische Physik, Universit\"at Regensburg, 93040 Regensburg, Germany}
}

\begin{abstract}
The quantum mechanical many-body problem is rarely analytically solvable. One notable exception is the case of two electrons interacting via a Coulomb potential in a uniform magnetic field. The motion is confined to a two-dimensional plane, which is commonly the case in nanodevices. We compare the exact solution with the semiclassical energy spectrum and study the time-dependent dynamics of the system using the time-dependent variational principle.
\end{abstract}

\maketitle


\section{Introduction}

The study of the motion of charged particles in an external magnetic field is of importance in many fields of physics. Astrophysical simulations need to model plasmas, while in solid-state physics at low temperatures magnetic fields cause the integer and fractional quantum Hall effects. In all cases the understanding of interactions is required to describe electronic transport since the motion of charged particles in an external magnetic field is driven by the electric fields generated by the charges themselves, and applied externally \cite{Kramer2009c,Kramer2009b}. Few-electron systems can be engineered with great precision in nanodevices in quantum-dots and the excitation spectrum of the system has been studied experimentally and theoretically \cite{Lipparini2008}. The presence of interactions causes intricate correlation effects, most prominently discussed within the fractional quantum Hall effect. But before one can proceed to large systems, it is important to reduce the complexity and study the simplest example, which still permits an analytical and numerical exact solution. For two- and many electron-atoms Moshinsky, Novaro, and Calles devised the pseudo-atom \cite{Moshinsky1968,Moshinsky1970a}, where the Coulomb interactions are replaced by a repelling harmonic oscillator potential in order to solve the many-body problem exactly and to obtain quantitative error bounds for the Hartree-Fock method.
Interacting electrons in a magnetic field show a particular rich behavior, since the repulsion leads to an initial increase in distance. This is  converted by the velocity-dependent Lorentz force into the cyclotron motion, which brings all charges back together and no particle escape is possible if the system is confined to two-dimensions.
The solution of Fock \cite{Fock1928}, Landau \cite{Landau1930}, and Darwin \cite{Darwin1931} for the limit of the non-interacting case are discussed in many textbooks on nanodevices \cite{Ferry1997a,Lipparini2008}, while the solution for the interacting case has been found many decades later \cite{Vercin1991,Taut1994}. Already the classical system reveals a complicated dynamics due to the interplay of the Lorentz force in the magnetic field and the electric field generated by the electron-electron repulsion \cite{Curilef1997}. In nanodevices, electronic properties are often probed in a two-dimensional layer perpendicular to the magnetic field direction \cite{Ferry1997a,Davies1998a} and thus a reduction of the 
three-dimensional problem to a quasi two-dimensional one is justified and needed to understand the experimental measurements.

Here, we analyze the interacting two-electron problem in the absence of an additional confining potential, since we are interested in the intrinsic dynamical properties of the system. After studying the exact quantum-mechanical stationary solution, which is available only for specific parameters, we present a semiclassical analysis covering all interaction values. Then we present a time-dependent, dynamical view of the electron dynamics in the frame-work of the time-dependent variational principle (TDVP) and compare the TDVP to the numerical solution. The two-body part of an interacting quantum-mechanical systems already displays profound differences to the classical dynamics caused by the antisymmetrization requirement for fermionic dynamics \cite{Feldmeier2000a}. Finally we present an outlook how to proceed from the present two-body analysis to more particles in a systematic fashion, which is required to study transport in nanodevices.

\section{The spectrum of two interacting electrons}

Ignoring the extension of the wave-function along the third dimension in the direction of the magnetic field, the Hamiltonian for the two electrons is given by
\begin{equation}
H_{\rm ee}=
 \frac{\mathbf{p}_1^2}{2m}
+\frac{\mathbf{p}_2^2}{2m}
+\frac{1}{2}m\omega_l^2\left(\mathbf{r}_1^2+\mathbf{r}_2^2\right)
-\omega_l (L_{z,1}+L_{z,2})
+\frac{\kappa}{|\mathbf{r}_1-\mathbf{r}_2|}
\end{equation}
with Larmor frequency $\omega_L=\frac{e B}{2m}$ and angular momentum 
$L_{z,i}=-\rmi\hbar(x_i \partial_{y,i}-y_i \partial_{x,i})$, $i=1,2$. The strength of the Coulomb interaction is given by the parameter $\kappa$.
The Hamiltonian $H_{\rm ee}$ separates into a center-of-mass term $H_{\rm cm}$ and a relative term $H_{\rm rel}$ in Jacobi coordinates
\begin{eqnarray}
H_{\rm cm}&=&
 \frac{\mathbf{P}^2}{2M}
+\frac{1}{2}M\omega_l^2 \mathbf{R}^2
-\omega_l L_Z,\\\label{eq:Hrel}
H_{\rm rel}&=&
 \frac{\mathbf{p}^2}{2\mu}
+\frac{1}{2}\mu\omega_l^2 \mathbf{r}^2
-\omega_l L_z
+\frac{\kappa}{|\mathbf{r}|},
\end{eqnarray}
with respective coordinates and masses
\begin{eqnarray}
\mathbf{R}=\frac{\mathbf{r}_1+\mathbf{r}_2}{2},\quad & \mathbf{r}=\mathbf{r}_2-\mathbf{r}_1,\\
\mathbf{P}=-\rmi\hbar\nabla_\mathbf{R},\quad & \mathbf{p}=-\rmi\hbar\nabla_\mathbf{r}, \\
M=2 m,\quad  & \mu=m/2.
\end{eqnarray}
Only the relative part of the Hamiltonian contains the Coulomb interaction, and the solutions for the non-interacting center-of-mass Hamiltonian $H_{\rm cm}$ are Landau levels with eigenfunctions and energies given by products of exponential functions and Hermite polynomials \cite{Gottfried2004a}. The ground state attains a gaussian shape with the eigenfunction and eigenenergy given by
\begin{equation}
\psi_{\rm cm}^{(0)}(X,Y)=\frac{1}{A\sqrt{\pi}}\exp\left(-\frac{X^2+Y^2}{2A^2}\right),\quad
A=\sqrt{\frac{\hbar}{M\omega_l}}, \quad
E_{\rm cm}^{(0)}=\hbar\omega_l.
\end{equation}

\subsection{Exact solutions for particular interaction strengths}

The solutions of the relative part $H_{\rm rel}$ are available in closed form for specific ratios of the Coulomb interaction and the magnetic field strength \cite{Vercin1991,Taut1994,Taut2009}. Only in these cases the eigenstates are given by the product of a polynomial of finite degree and a decaying exponential function. However we do not view the absence of finite-degree polynomial solutions as an indication for the appearance of anyons (as done in \cite{Vercin1991}), but find numerically and semiclassicaly normalizable solutions for all values of the interaction strength. The conditions for a closed form solution are related to a hidden symmetry within the problem \cite{Turbiner1994a,Tempesta2001}. Here, we list the first solutions classified by the angular momentum $l$ and the quantum number $n_r$
\begin{equation}
\psi_{\rm rel}^{(n_r,l)}(x,y)=N^{(n_r,l)}\;R^{(n_r)}(r)\rme^{\rmi\theta l},
\, r=\sqrt{x^2+y^2},
\, \theta=\arctan(x,y),
\, l=0,\pm 1,\pm 2,\ldots
\end{equation}
up to the normalization factor $N^{(n_r,l)}$. We set $\hbar=\mu=\omega_l=1$ in Eq.~\ref{eq:Hrel} and thus all energies are given in units of $\hbar\omega_l$, all lengths in units of $2\sqrt{\frac{\hbar}{eB}}$, and the Coulomb interaction strength in units of $\frac{\hbar^2}{m}\sqrt{\frac{eB}{\hbar}}$.\\
For $\kappa=\sqrt{|l|+1/2}$
\begin{eqnarray}
\psi^{(1,l)}_{\rm rel}(r,\theta)=&N^{(1,l)} r^{|l|} \left[1+r/\sqrt{|l|+1/2}\right] \rme^{-r^2/2} \rme^{\rmi\theta l} ,\\
E^{(1,l_-)}=&2+2|l|  &\text{for} \quad l<0 \nonumber ,\\
E^{(1,l_+)}=&2       &\text{for} \quad l>0 \nonumber .
\end{eqnarray}
For $\kappa=\sqrt{4|l|+3}$
\begin{eqnarray}
\psi^{(2,l)}_{\rm rel}(r,\theta)=&N^{(2,l)} r^{|l|} \left[1+2|l|+2 r (\sqrt{4|l|+3}+r)\right] \rme^{-r^2/2} \rme^{\rmi\theta l},\\
E^{(2,l_-)}=&3+2|l| & \text{for} \quad l<0 \nonumber ,\\
E^{(2,l_+)}=&3      & \text{for} \quad l>0 \nonumber .
\end{eqnarray}
For $\kappa=\sqrt{5+5|l|\pm\frac{1}{2}\sqrt{73+64|l|(|l|+2)}}$
\begin{eqnarray}
\psi^{(3,l)}_{\rm rel}(r,\theta)=&N^{(3,l)} r^{|l|} \left[1+\frac{2r\kappa}{2|l|+1}-\frac{8r^2\kappa(\kappa+r)}{(2|l|+1)(9+6|l|-2\kappa^2)} \right] \rme^{-r^2/2} \rme^{\rmi\theta l},\\
E^{(3,l_-)}=&4+2|l| & \text{for} \quad l<0 \nonumber ,\\
E^{(3,l_+)}=&4      & \text{for} \quad l>0 \nonumber .
\end{eqnarray}

\subsection{Semiclassical WKB spectrum}

\begin{figure}[t]
\includegraphics[width=0.99\textwidth]{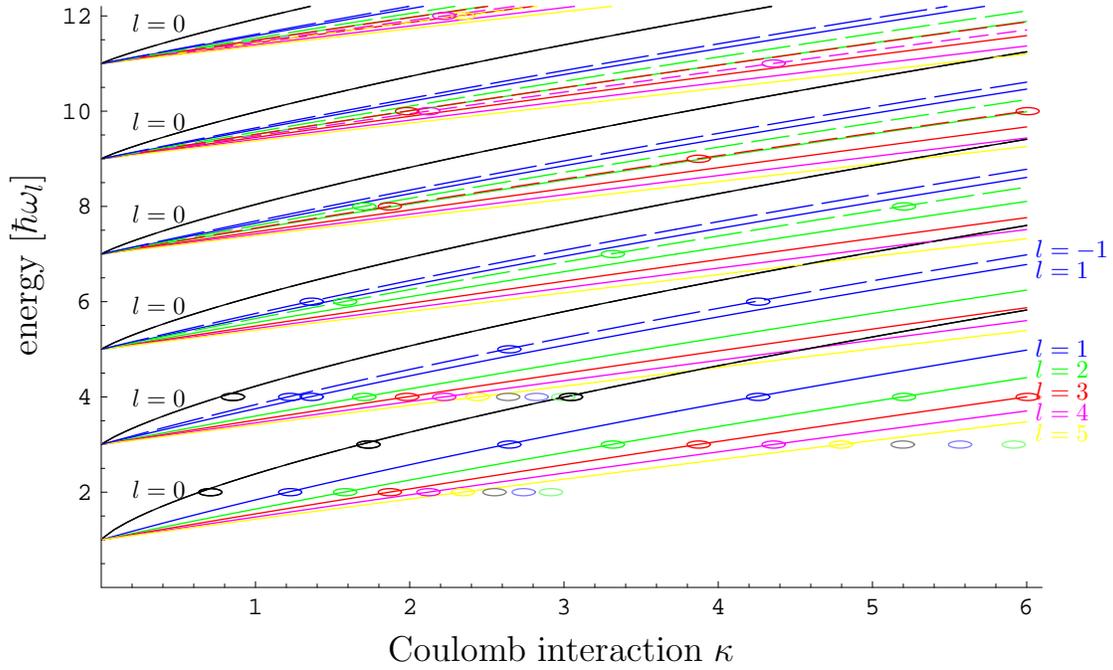}
\caption{\label{fig:SCspectrum}(Color online) Semiclassical spectrum $E(\kappa)$. Color denotes the value of the angular momentum $l$ (solid lines $l\ge0$, dashed lines $l<0$, black: $l=0$, blue: $l=\pm 1$, green $l=\pm 2$, red $l=\pm 3$, magenta $l=\pm 4$, yellow $l= \pm5$. The ellipsoids denote exact quantum-mechanical solutions based on Heun polynomials.}
\end{figure}

\begin{figure}[t]
\includegraphics[width=0.99\textwidth]{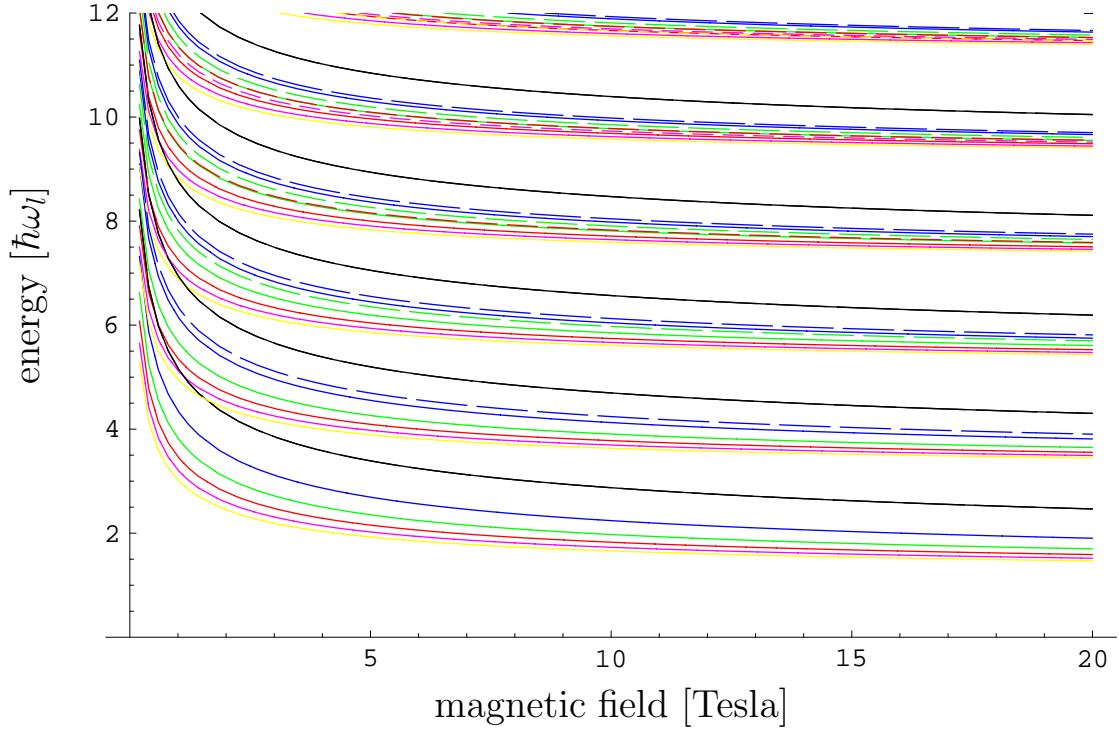}
\caption{\label{fig:SCspectrumB}(Color online) Semiclassical spectrum $E(B)/(\hbar\omega_l)$ as function of magnetic field $B$ in Tesla, (effective electron mass $m^*=0.1$, dielectric constant $\epsilon=10$). Color denotes the value of the angular momentum $l$ (solid lines $l\ge0$, dashed lines $l<0$, black: $l=0$, blue: $l=\pm 1$, green $l=\pm 2$, red $l=\pm 3$, magenta $l=\pm 4$, yellow $l= \pm5$.}
\end{figure}

To construct the entire energy spectrum at interaction values between the ones tabulated above requires numerical methods or alternatively to employ approximation schemes. The semiclassical energy spectrum can be obtained for all interaction strengths by following Ref.~\cite{Klama1998a}, where we again set the external confinement to zero. The ansatz
\begin{equation}
\psi^{(n,l)}(r,\theta)=\rme^{\rmi l \theta}\frac{\chi(r)}{\sqrt{r}}
\end{equation}
leads to the radial equation
\begin{equation}
\chi''(r)+\left[2E+2l-\frac{l^2-1/4}{r^2}-\frac{2\kappa}{r}-r^2\right]\chi(r)=0.
\end{equation}
Upon making the Langer correction appropriate for two-dimensions, $l^2-1/4\rightarrow l^2$ \cite{Berry1972a}, the semiclassical quantization condition reads
\begin{equation}\label{eq:RadialEquation}
\int_{r_1}^{r_2}\rmd r \sqrt{2E+2l-\frac{l^2}{r^2}-\frac{2\kappa}{r}-r^2}=\pi(n+1/2),
\end{equation}
where $r_{1,2}$ denote the inner and outer turning points (zeroes) of the radial equation. The polynomial in the square root has four zeroes,
$a$, $b$, $c$ and $d$, which are ordered such that $b>a$ and $c,d<0$. For given $\kappa$, $n$ and $m$ we obtain an implicit equation for the energies $E^{(n,l)}$. Once $a,b,c,d$ are known, the integral~(\ref{eq:RadialEquation}) is given in Ref.~\cite{Klama1998a}
\begin{equation}
-\frac{3}{2}\kappa I_0+(E+l)I_1-l^2 I_2=\pi(n+1/2),
\end{equation}
and written in terms of the complete elliptic integrals $K$, $\Pi$
\begin{eqnarray}
I_0&=&\frac{2}{\sqrt{(a-c)(b-d)}}K\left(k\right)\\
I_1&=&\frac{2}{\sqrt{(a-c)(b-d)}}\left[(b-c)\Pi\left(\frac{a-b}{a-c}\bigg|k\right)+c K\left(k\right)\right]\\
I_2&=&\frac{2}{\sqrt{(a-c)(b-d)}}\left[\frac{c-b}{bc}\Pi\left(\frac{c(a-b)}{b(a-c)}\bigg|k\right)+\frac{1}{c}K\left(k\right)\right]\\
k^2&=&\frac{(a-b)(c-d)}{(a-c)(b-d)}.
\end{eqnarray}
The semiclassical spectrum as a function of $\kappa$ is shown in Fig~\ref{fig:SCspectrum}. The ellipsoids mark exact results from the quantum solution at specific values of $\kappa$. The dashed lines denote a reversed sense of rotation ($l<0$) compared to the direction preferred by the Lorentz force. The states with a reversed sense of rotation $l<0$ are shifted to higher energies. Our results are in line with the Landau level degeneracy for the non-interacting case, where one obtains \cite{Gottfried2004a}, p.~193 that the Landau level of quantum number $n$ ($n=0,1,2,\ldots,\infty)$ contains the angular momenta $l=-n,-n+1,-n+2,\ldots,\infty$. The Coulomb interaction lifts this degeneracy and reveals the number of negative angular momenta states contained in each Landau level. Note that the Coulomb interaction $\kappa$ depends in SI-units on the magnetic field. The spectrum for fixed Coulomb interaction as a function of magnetic field is show in Fig.~\ref{fig:SCspectrumB}.

\section{Time-dependent solution}

In order to understand the role of interactions and antisymmetrization in scattering effects between two electrons, it is instructive to consider the time-dependent dynamics of two initially separated electrons. For simplicity we restrict ourselves to cases where the center-of-mass part of the system remains in the ground state and we only consider the time-evolution of the relative part. As initial orbitals for the two electrons, we pick two shifted Gaussian functions which provide an exact solution in the absence of the Coulomb interaction
\begin{eqnarray}
\phi^a(\mathbf{r})&=&\frac{1}{a\sqrt{\pi}}\exp\left(-\frac{{(x-x_o)}^2+y^2}{2a^2}\right),\\
\phi^b(\mathbf{r})&=&\frac{1}{a\sqrt{\pi}}\exp\left(-\frac{{(x+x_o)}^2+y^2}{2a^2}\right).
\end{eqnarray}
From the two orbitals we form the symmetric and antisymmetric and not-symmetrized combinations for the two-electron wave function
\begin{eqnarray}
\psi^A(\mathbf{r}_1,\mathbf{r}_2)&=&[\phi^a(\mathbf{r}_1)\phi^b(\mathbf{r}_2)-\phi^a(\mathbf{r}_2)\phi^b(\mathbf{r}_1)]/\sqrt{N^A}\\
\psi^B(\mathbf{r}_1,\mathbf{r}_2)&=&[\phi^a(\mathbf{r}_1)\phi^b(\mathbf{r}_2)+\phi^a(\mathbf{r}_2)\phi^b(\mathbf{r}_1)]/\sqrt{N^B}\\
\psi^C(\mathbf{r}_1,\mathbf{r}_2)&=&[\phi^a(\mathbf{r}_1)\phi^b(\mathbf{r}_2)].
\end{eqnarray}
Since the orbitals are not orthogonal to each other, the normalization $N$ differs from $\sqrt{2}$. The width parameter $a$ of the Gaussian is chosen such that the electrons are in the lowest Landau level $a=\sqrt{\frac{2\hbar}{e B}}$. We place the two electrons initially at rest (this requires a gauge factor!) at a separation of $d=2 x_o$. Next we transform the states to relative and center-of-mass coordinates, where the wavefunction becomes a product
\begin{equation}
\psi(\mathbf{r}_1,\mathbf{r}_2)=\psi_{\rm rel}(\mathbf{r}_2-\mathbf{r}_1)\psi_{\rm cm}((\mathbf{r}_1+\mathbf{r}_2)/2).
\end{equation}
In relatives coordinates we obtain ($\mu=\hbar=\omega_l=1$)
\begin{eqnarray}
\psi_{\rm rel}^A(\mathbf{r})&=&\sqrt{\frac{2}{\pi(\rme^{x_o^2}-1)}}\sinh(x x_o)\rme^{-(x^2+y^2)/2},\\
\psi_{\rm rel}^B(\mathbf{r})&=&\sqrt{\frac{2}{\pi(\rme^{x_o^2}+1)}}\cosh(x x_o)\rme^{-(x^2+y^2)/2},\\
\psi_{\rm rel}^C(\mathbf{r})&=&\sqrt{\frac{2}{\pi}}\rme^{-({(x-x_o)}^2+y^2)/2},
\end{eqnarray}
and the center-of-mass wavefunction reads
\begin{equation}
\psi_{\rm cm}^{(0)}(X,Y)=\frac{2}{\sqrt{\pi}}\rme^{-2 (X^2+Y^2)}.
\end{equation}
The total wave-function is then recovered via
\begin{equation}
\psi^{A,B,C}(x_1,y_1;x_2,y_2)=\psi_{\rm cm}^{(0)}((x_1+x_2)/2,(y_1+y_2)/2)\psi_{\rm rel}^{A,B,C}(x_2-x_1,y_2-y_1).
\end{equation}
As time-dependent observable quantities we track the density of the relative motion
\begin{equation}
\rho_{\rm rel}(\mathbf{r};t)={|\psi_{\rm rel}(\mathbf{r};t)|}^2
\end{equation}
and the single-particle density obtained by integrating over one set of particle coordinates
\begin{eqnarray}
\rho_{\rm spd}(\mathbf{r}_1;t)
&=&\int\rmd\mathbf{r}_2\;{|\psi(\mathbf{r}_1,\mathbf{r}_2)|}^2\\
&=&\int\rmd\mathbf{r}_2\;
{|\psi_{\rm rel}(\mathbf{r}_2-\mathbf{r}_1;t)|}^2
{|\psi_{\rm cm}((\mathbf{r}_1+\mathbf{r}_2)/2;t)|}^2
\end{eqnarray}
The single-particle densities of the initial states are
\begin{eqnarray}
\rho_{\rm spd}^A(\mathbf{r})&=&\frac{2\rme^{-2(x^2+y^2)}}{\pi(\rme^{x_o^2}-1)}\left(\rme^{x_o^2/2}\cosh(2x_o x)-1\right)\\
\rho_{\rm spd}^B(\mathbf{r})&=&\frac{2\rme^{-2(x^2+y^2)}}{\pi(\rme^{x_o^2}+1)}\left(\rme^{x_o^2/2}\cosh(2x_o x)+1\right)\\
\rho_{\rm spd}^C(\mathbf{r})&=&\frac{2\rme^{-2({(x-x_o/2)}^2+y^2)}}{\pi},
\end{eqnarray}
with normalization
\begin{equation}
\int\rmd\mathbf{r}\;\rho_{\rm spd}^{A,B,C}(\mathbf{r})=1.
\end{equation}

\subsection{Numerical time-propagation}

The initial state in the relative coordinate can be represented in discretized form on a regular two-dimensional mesh, and the time evolution is performed using a split-operator method adopted to the presence of a magnetic field \cite{Kramer2008a}. The Hamiltonian is represented in a mixed momentum-position representation necessary due to the angular-momentum operator $L_z$. For small time-steps $\Delta t$ the time-evolution is given by 
\begin{eqnarray}
\psi(\mathbf{r},t'+N\Delta t)&\approx&
\rme^{-\rmi\Delta t/\hbar\;V(x,y)/2}
{\left[
{\cal F}^{-1}_y
\rme^{-\rmi\Delta t/\hbar\;T_{py,x}}
{\cal F}_y
{\cal F}^{-1}_x
\rme^{-\rmi\Delta t/\hbar\;T_{px,y}}
{\cal F}_x
\rme^{-\rmi\Delta t/\hbar\;V(x,y)}
\right]}^N\nonumber\\
&&\quad\times\rme^{\rmi\Delta t/\hbar\;V(x,y)/2}\psi(\mathbf{r},t'),
\end{eqnarray}
where ${\cal F}_x, {\cal F}_y$ denote partial Fourier transforms with respect to only one-dimension and
\begin{equation}
T_{p_x,y}=\frac{p_x^2}{2\mu}-\omega_L p_x y,\quad
T_{p_y,x}=\frac{p_y^2}{2\mu}+\omega_L p_y x,\quad
V(x,y)=\frac{1}{2}\mu\omega_l^2\mathbf{r}^2+\frac{\kappa}{|\mathbf{r}|}.
\end{equation}
The coherent-state propagator for vanishing Coulomb interaction is known in closed form \cite{Dodonov1975} and can be used to test the numerical scheme.

\subsection{Time-dependent variational principle}

Next we consider the dynamics of two electrons based on the application of the time-dependent variational principle (TDVP) due to Dirac and Frenkel, which has been reviewed in \cite{Kramer1981a,Ohrn2005a}. The starting point is the action
\begin{equation}
S=\int_{t_1}^{t_2}\rmd t\; L(\psi^*,\psi),\quad L=\frac{\langle\psi|\rmi\partial_t-H|\psi\rangle}{\langle\psi|\psi\rangle}.
\end{equation}
The principle of extremal action requires $\delta S=\int\rmd t\;\delta L=0$. For the Gaussian orbitals used in the previous section is is convenient to use a notation related to coherent states and we label the parameters of the variational wavefunction by $z$: $\psi=\psi(z)=|z\rangle$.
Using $i\partial_t=\frac{\rmi}{2}(-\overleftarrow{\partial_t}+\overrightarrow{\partial_t})$ we get
\begin{equation}
\begin{split}
\delta L
&=\frac{\frac{\rmi}{2}\langle\delta z|\dot{z}\rangle-\frac{\rmi}{2}\langle\delta \dot{z}|z\rangle-\langle z|H|z\rangle}{\langle z|z\rangle}\\
&-\left[\frac{\rmi}{2}\langle z|\dot{z}\rangle-\frac{\rmi}{2}\langle \dot{z}|z\rangle-\langle z|H|z\rangle\right]\frac{\langle\delta z|z\rangle}{{\langle z|z\rangle}^2}+\text{c.c.}
\end{split}
\end{equation}
The time-derivatives of the form $\delta\dot{z}$ are removed using the identity
\begin{equation}
\frac{\rmd}{\rmd t}\frac{\langle\delta z|z\rangle}{\langle z|z\rangle}=\frac{\langle \delta\dot{z}|z\rangle+\langle\delta z|\dot{z}\rangle}{\langle z|z\rangle}
-\frac{\langle\delta z|z\rangle}{{\langle z|z\rangle}^2}\frac{\rmd}{\rmd t}\langle z|z\rangle
\end{equation}
and we rewrite $\delta L$
\begin{equation}
\delta L=
\rmi\frac{\langle\delta z|\dot{z}\rangle}{\langle z|z\rangle}
-\frac{\langle \delta z|H|z\rangle}{\langle z|z\rangle}
-\frac{\langle\delta z|z\rangle}{{\langle z|z\rangle}^2}\big[\rmi\langle z|\dot{z}\rangle-\langle z|H|z \rangle\big]+\text{c.c.},
\end{equation}
where we dropped all total time derivatives since the variation vanishes at $t_1$ and $t_2$. Finally we obtain from the variation with respect to $\delta z$
\begin{equation}
\left(\rmi\partial_t-H\right)|z\rangle=\frac{\langle z|\rmi\partial_t-H|z\rangle}{\langle z|z\rangle}|z\rangle,
\end{equation}
the equations of motion for the parameterized wave function by denoting
\begin{equation}
|\psi\rangle=|\psi(z_1\dots z_N)\rangle,\quad 
N=\langle\psi(z)|\psi(z)\rangle,\quad
{\cal H}=\frac{\langle\psi(z)|H|\psi(z)\rangle}{\langle\psi(z)|\psi(z)\rangle}
\end{equation}
and introducing $\delta z=\sum_i\partial_{z_i}\delta z_i$
\begin{equation}
\delta L=
 \sum_i\left[\sum_j \rmi\frac{\ln N}{\partial z_i^*\partial z_j}\dot{z}_j  -\partial_{z^*_i} {\cal H}\right]\delta z^*_i
+\sum_i\left[\sum_j-\rmi\frac{\ln N}{\partial z_i\partial z_j^*}\dot{z}_j^*-\partial_{z_i  } {\cal H}\right]\delta z^*_i.
\end{equation}
For a compact notation we introduce the matrix 
\begin{equation}
C_{ij}=\frac{\partial^2}{\partial{z_i}\partial{z_j^*}}\ln N,
\end{equation}
which allows us to recast the variational equations in a shape familiar from Hamilton's equations
\begin{equation}
\left(
\begin{array}{cc}
-\rmi C^{-1} & 0 \\
0 & \rmi {C^*}^{-1}
\end{array}
\right)
\left(
\begin{array}{l}
\partial_{z^*} {\cal H}\\
\partial_{z} {\cal H}
\end{array}
\right)
=
\left(
\begin{array}{l}
\dot{z}\\
\dot{z}^*
\end{array}\right).
\end{equation}
In the ansatz for the variational function the phase and normalization were explicitly removed. A normalized state with phase is restored by multiplying the variational state $\psi$ by an exponential factor
\begin{equation}
|\phi\rangle=|\psi(z)\rangle\exp\left[\rmi\int^t\rmd t' \frac{\langle \psi(z)|\rmi \partial_{t'}-H|\psi(z)\rangle}{\langle \psi(z)|\psi(z)\rangle}\right].
\end{equation}

\subsection{Choice of variational wave-function}

\begin{figure}[t]
\begin{minipage}{0.85\textwidth}
\begin{center}
\includegraphics[width=0.85\textwidth]{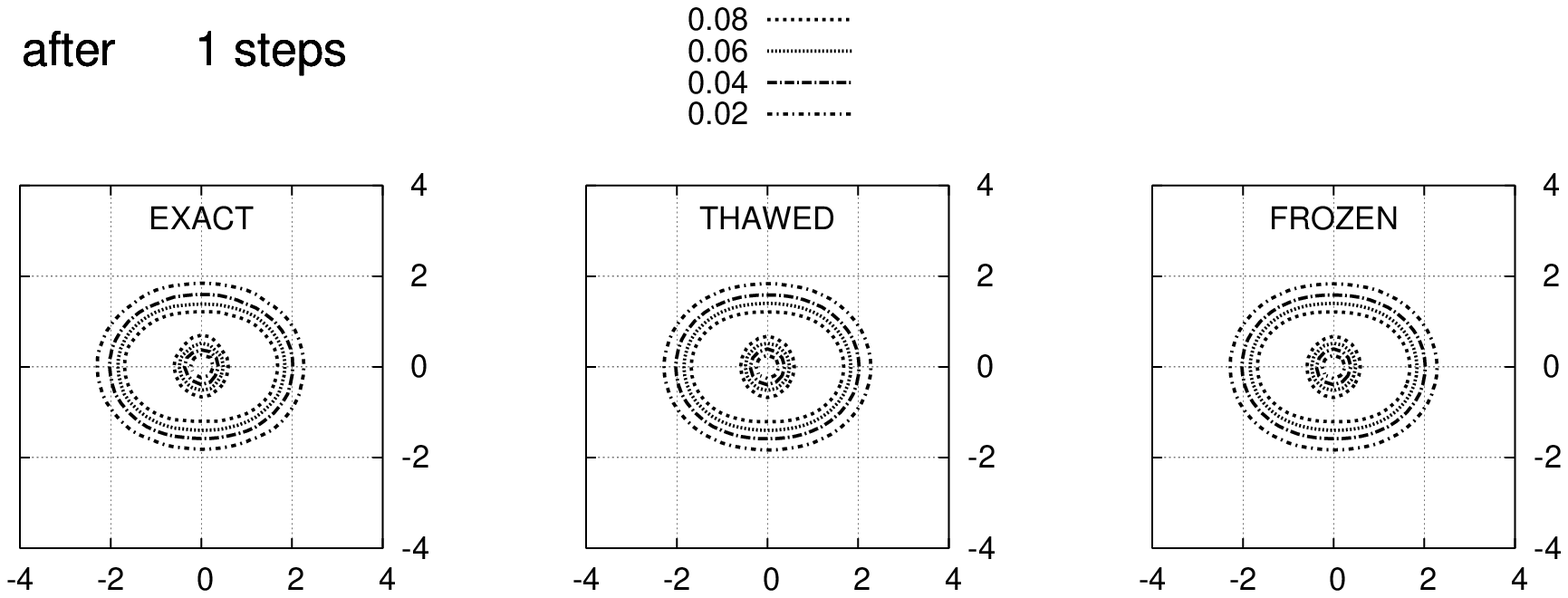}\\
\includegraphics[width=0.85\textwidth]{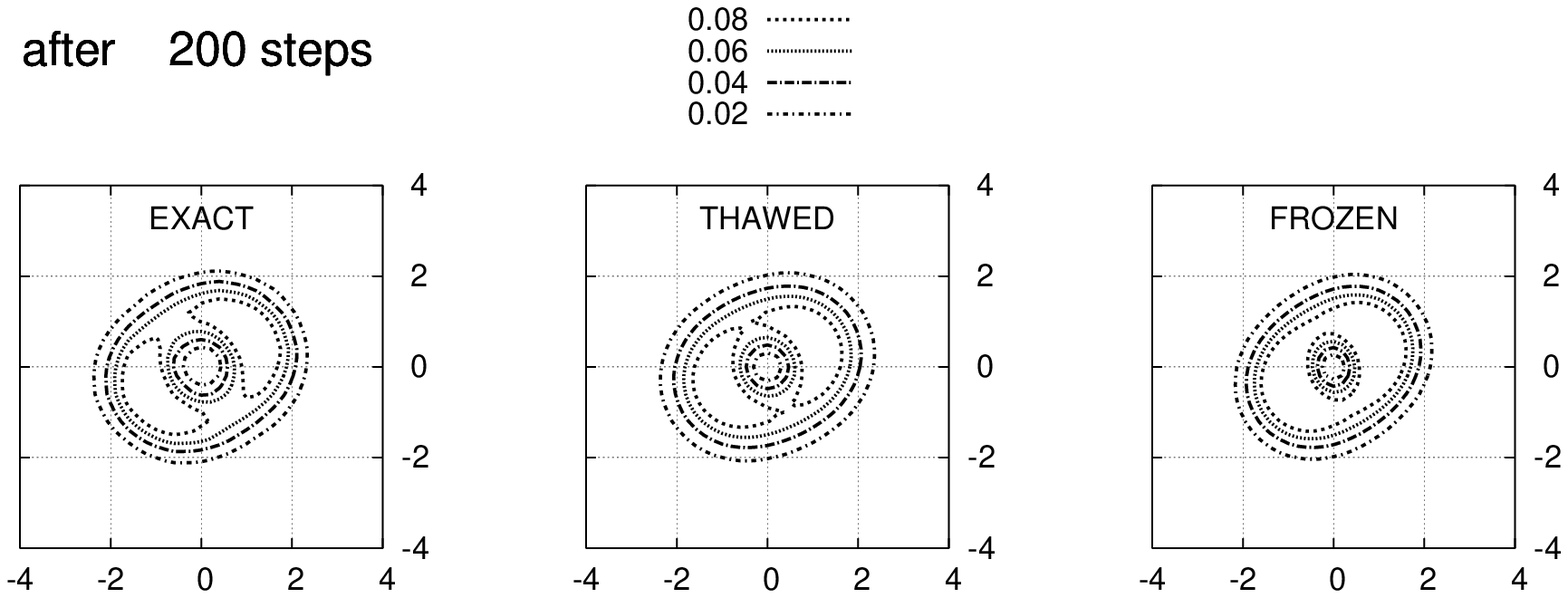}\\
\includegraphics[width=0.85\textwidth]{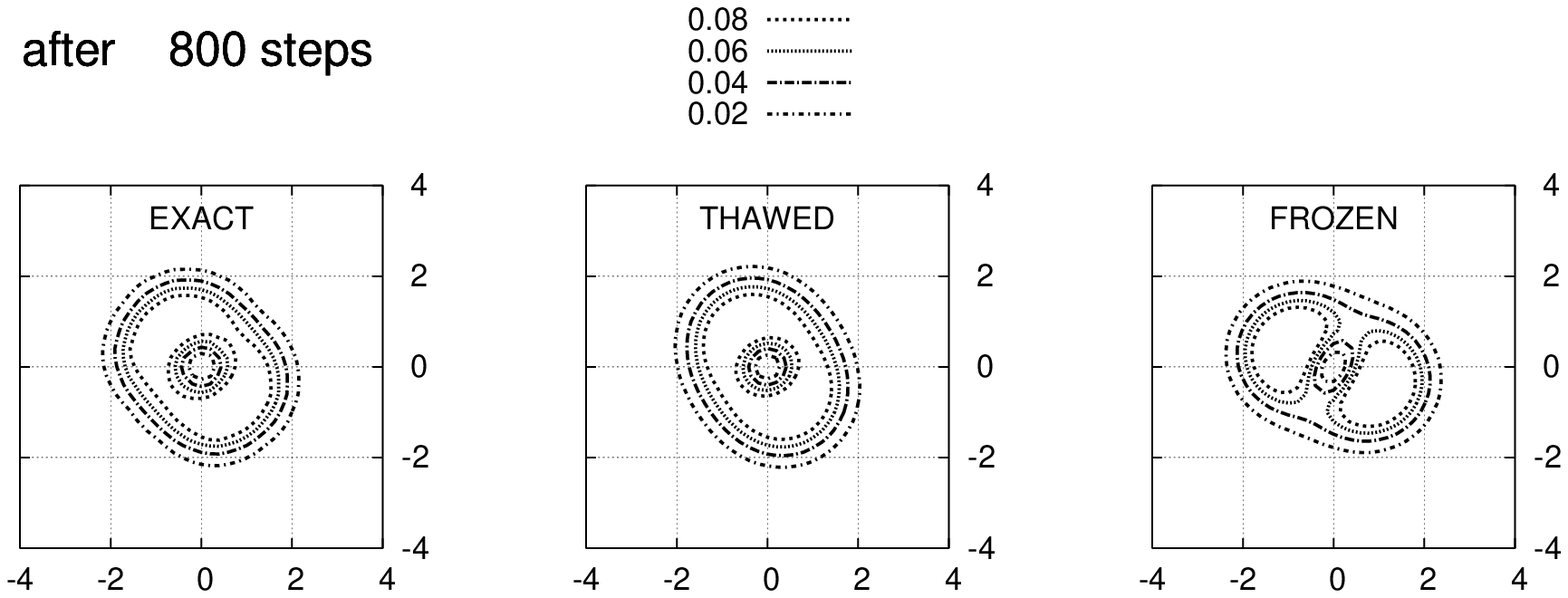}
\end{center}
\end{minipage}
\caption{\label{fig:timeevo}Comparison of the time-evolved densities ${|\psi_{\rm rel}(x,y)|}^2$ after a different number of propagation steps (time step $\Delta t=0.02$, Coulomb interaction $\kappa=1$, initial offset $x_o=1/\sqrt{2}$, initial width $w=1/\sqrt{2}$). The left column shows the exact result, the middle column the ``thawed'' wavepacket of a squeezed coherent state with all six variational degrees of freedom, and the right column the fixed-width (``frozen'') TDVP result for a coherent state. With increasing time the TDVP solutions start to deviate from the exact solution.}
\end{figure}

Any variational principle can return only good results if the variational functions are chosen with physical insight into the problem at hand. The choice of coherent states for electron-orbitals in a magnetic field is physically motivated by their property to restore the exact result for the non-interacting case. The Coulomb interaction distorts the coherent state evolution by changing the trajectories and momenta encoded in $q_x,q_y,p_x,p_y$. In addition the width of the wave-function is no-longer fixed and we generalize our ``frozen'' coherent-state variational functions to include squeezed coherent states with adaptable ``thawed'' widths. A variable width requires to consider two more variables, since the time-evolution leads to a dispersion which has to be represented by a complex-valued parameter \cite{Feldmeier2000a}. In total we describe the relative motion by six variational parameters $q_x,q_y,p_x,p_y,u,w$ based on the function
\begin{equation}
\psi^{C}_{\rm rel}(x,y;p_x,p_y,q_x,q_y,u,w)=\rme^{-\frac{(1-2\rmi uw)}{4w^2}({{(q_x-x)}^2+{(q_y-y)}^2})+\rmi p_x x+\rmi p_y y},
\end{equation}
which we antisymmetrize
\begin{eqnarray}
\psi^{A}_{\rm rel}(x,y;p_x,p_y,q_x,q_y,u,w)&=&\psi^{C}_{\rm rel}(x,y;p_x,p_y,q_x,q_y,u,w)\notag\\&&-\psi^{C}_{\rm rel}(x,y;-p_x,-p_y,-q_x,-q_y,-u,-w)
\end{eqnarray}
in order to describe two electrons with the spin pointing in the same direction. It is instructive to compare the expectation value of the Hamiltonian with respect to $\psi^{C}_{\rm rel}$ and $\psi^{A}_{\rm rel}$. In the first case we obtain
\begin{eqnarray}
{\cal H}^C
&=&\frac{{}^C\langle \psi | H_{\rm rel} | \psi \rangle^C}{{}^C\langle \psi |\psi \rangle^C}\notag\\
&=&
\underbrace{\frac{p_x^2+p_y^2}{2\mu}
+\frac{1}{2}\mu\omega_l^2 q^2
-\omega_l (p_y q_x - p_x q_y)
+\kappa\frac{\sqrt{\pi}\rme^{-q^2/(4w^2)}}{\sqrt{2}w}\BesselI_0(q^2/(4w^2))}_{\text{``classical''}}\notag\\
&&+\underbrace{\frac{u^2}{\mu}
+\frac{1}{4 w^2\mu} 
+w^2 \mu \omega_l^2}_{\text{``uncertainty''}},
\end{eqnarray}
where the the first parts corresponds to the Hamiltonian of the classical charge density ${|\psi_{\rm rel}|}^2$ and the last part adds the minimum uncertainty and represents a quantum-mechanical effect. The antisymmetrized expectation value contains in addition correlation and exchange terms
\begin{eqnarray}
{\cal H}^A
&=&\frac{{}^A\langle \psi | H_{\rm rel} | \psi \rangle^A}{{}^A\langle \psi |\psi \rangle^A}\\
&=&
\frac{p_x^2+p_y^2}{2\mu}
+\frac{1}{2}\mu\omega_l^2 q^2
-\omega_l (p_y q_x - p_x q_y)
+\frac{1}{4 w^2\mu} 
+w^2 \mu \omega_l^2\notag\\
&&
+\kappa\left(1+\frac{1}{\rme^\xi-1}\right)
\bigg[\frac{\sqrt{\pi}\rme^{-q^2/(4w^2)}}{\sqrt{2}w}\BesselI_0(q^2/(4w^2))\notag\\
&&\quad-\frac{\sqrt{\pi}\rme^{-\xi/2+q^2/(4 w^2)}}{\sqrt{2}w}\BesselI_0\left((q_x u-p_x w)^2+(q_y u-p_y w)^2\right)
\bigg]\notag
\end{eqnarray}
where $\xi$ denotes a distance in phase space
\begin{eqnarray}
\xi&=&2\left({(u q_x-w p_x)}^2+{(u q_y-w p_y)}^2\right)+\frac{q_x^2+q_y^2}{2w^2}.
\end{eqnarray}
For large values of $\xi$ the antisymmetrized expectation value of the Hamiltonian approaches ${\cal H}^C$. In order to complete the derivation of the equations of motion, the matrix $C$ has to be determined. The resulting expressions are lengthy and not reproduced here. For the antisymmetrized case, all variational parameters are now coupled, whereas for the unsymmetrized matrix a fully sympletic relation holds for the position and momenta as well as for the width parameters
\begin{equation}
C=
\left(\begin{array}{cccccc}
0 & 0 & 1 & 0 & 0 & 0\\
0 & 0 & 0 & 1 & 0 & 0\\
-1 & 0 & 0 & 0 & 0 & 0\\
0 & -1 & 0 & 0 & 0 & 0\\
0 & 0 & 0 & 0 & 0 & 2\\
0 & 0 & 0 & 0 & -2 & 0
\end{array}\right).
\end{equation}
As for the expectation value of the Hamiltonian, the antisymmetrized $C$ matrix approaches the symplectic form for large distances in phase space.
The system of coupled differential equations for the time-evolution of the initial state is now completely determined
\begin{equation*}
\mathbf{C}
\left(\begin{array}{c}
\dot{q_x}\\
\dot{q_y}\\
\dot{p_x}\\
\dot{p_y}\\
\dot{u} \\
\dot{w} 
\end{array}\right)
=
\left(\begin{array}{c}
\partial_{q_x} {\cal H}\\
\partial_{q_y} {\cal H}\\
\partial_{p_x} {\cal H}\\
\partial_{p_y} {\cal H}\\
\partial_u {\cal H}\\
\partial_v {\cal H}
\end{array}\right),
\end{equation*}
but the integration has to be carried out numerically. We compare two different cases to the exact propagation in order to determine the quality of the variational function. The first case restricts the variational degrees of freedom to only four parameters, by keeping $w$ fixed and setting $u=0$, while the second approach includes all six parameters. The two TDVP solutions are compared to the numerically exact solution which we obtain with the wave-packet propagation scheme described above. In Fig.~\ref{fig:timeevo} we show the density of the relative wave-function for different propagation times. After starting from the same initial state, the TDVP solutions deviate with longer times. The frequency of rotation around the origin is a sensitive measure of the shape of the wave-function, since a slight change in the width of the wavefunction alters the electric field and affects the Lorentz-force induced rotation. A mechanical analog is the change of rotation rate of a spinning object due to changes of its moment of inertia. The thawed (squeezed) coherent-state can adapt its rotation rate better than the frozen-width coherent-state and thus stays closer to the exact solution for a longer time-evolution.

\section{Outlook: from two to many electrons}

The two-electrons is a special case of the general many-body problem. The addition of more particles complicates the problem considerably and no exact analytic solutions are known for the general case. In the case of a classical system of interacting particles, the self-generated fields of the electronic charges are the foundation of the classical Hall effect. We have demonstrated in \cite{Kramer2009c} that realistic boundary conditions and the Coulomb interactions completely alter the current-density distribution and produce the Hall field, in contrast to the non-interacting solution. The inclusion of quantum-effects in a classical calculation requires to account for the Pauli-principle and to recover the Fermi-Dirac statistics of fermionic systems. The TDVP offers a possibility to systematically move in this direction \cite{Feldmeier2000a} which we are currently exploring.

\section*{Acknowledgments}

I thank the organizers of the XL Latin American School of Physics for their kind invitation to Mexico and the organizers of the symposium for their invitation to present this contribution dedicated to Marcos Moshinsky, a unforgettable teacher of physics and friend for decades. Additional financial support from the Emmy-Noether program of the Deutsche Forschungsgemeinschaft (grant KR 2889/2) is gratefully acknowledged. I thank Peter Kramer for helpful discussions and comments.

\bibliographystyle{unsrt}
\providecommand{\url}[1]{#1}

\end{document}